\documentclass[aps,twocolumn,prl,showpacs]{revtex4}

\usepackage{graphicx}
\usepackage{amssymb,amsmath}

\begin{document}

\title{Spontaneous flow states in active nematics: a unified picture}

\author{S. A. Edwards and J. M. Yeomans}

\affiliation{                    
   The Rudolf Peierls Centre for Theoretical Physics, 1 Keble Road, Oxford, OX1 3NP, UK
}

\begin{abstract}
Continuum hydrodynamic models of active liquid crystals have been used to describe dynamic self-organising 
systems such as bacterial swarms and cytoskeletal gels. A key prediction of such models is the existence of 
self-stabilising kink states that spontaneously generate fluid flow in quasi-one dimensional channels 
\cite{VoiEPL05}. 
Using simple stability arguments and numerical calculations we extend previous studies to give a complete 
characterisation of the phase space for both contractile and extensile particles (ie pullers and pushers) moving in 
a 
narrow channel as a function of their flow alignment properties and initial orientation. This gives a framework 
for unifying many of the results in the literature. We describe the response of the kink states to an imposed 
shear, and investigate how allowing the system to be polar modifies its dynamical behaviour.
\end{abstract}

\maketitle

Suspensions of flagellate bacteria swim in quasi-turbulent swirls that suggest long-range collective ordering 
\cite{DomPRL04}. Cytoskeletal filaments and motor-proteins spontaneously form stars and spirals without any 
external guide \cite{NedNature97}. These are just two examples of a class of self-organizing system that poses a 
novel challenge to physicists, as the ordering is intrinsically non-equilibrium in nature. The pattern formation is 
a dynamic phenomenon that relies on a continuous expenditure of energy by the individual particles as they actively 
generate forces on each other and/or the surrounding medium \cite{NedCurrOpin03}.

\begin{figure}
\includegraphics[width=3in]{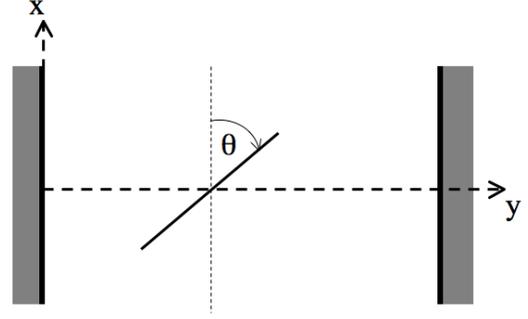}
\caption{Quasi-one dimensional slab geometry, with solid, no-slip walls lying parallel to the $x$-axis ensuring 
that only the $x$-component of the fluid velocity is non-zero. The orientation angle of the director field $\theta$ 
is measured such that $\theta = 0$ is parallel to the walls, and $\theta = \pi/2$ is perpendicular.}
\label{slab}
\end{figure}

Convergent strands of research focused on modeling bacterial swarms  \cite{SimPRL02, HatPRL04, AraPRE07,SaiPRL08} 
and on cytoskeletal dynamics \cite{LivPRL03, LivPRL06, KruPRL04, KruEPJE05} have suggested that the 
collective behaviour seen in both types of system can be described in the continuum limit by the same 
phenomenological, hydrodynamic model. This builds on the equations of liquid crystal hydrodynamics, 
which capture the inherent directionality of the particles (e.g. bacteria, microtubules), and adds 
extra non-equilibrium terms to account for the activity. One of the most striking predictions of the 
active liquid crystal model has been that, under certain conditions, the uniform aligned state is 
unstable \cite{SimPRL02}; furthermore, when confined to a quasi-one dimensional slab geometry by flat 
interfaces (fig.\  \ref{slab}) this instability can lead to the onset of a spontaneous steady 
flow \cite{VoiEPL05,VoiPRL06}. Previous studies of the spontaneous flow transition in the active liquid 
crystal model using both analytic \cite{VoiEPL05} and numerical \cite{MarPRL07,MarPRE07,MarJNNFM08} approaches 
have focused on illuminating only certain neighbourhoods of the complete phase space. To help map the 
phenomenological coefficients in the continnum theory onto microscopic parameters, and to aid in 
identifying their values for a given physical system, it is helpful to have a more complete picture of the 
effect of varying key coefficients. Therefore our aim in the Letter is to extend and unify 
previous studies by carrying out a complete numerical exploration of the phase space formed by 
three of the most important physical characteristics of the system: the type and strength of the flow 
field generated by the individual particles, their flow alignment behaviour, and the initial orientation 
of the director field. We distinguish six spontaneous flow states, describe their response to an applied shear, and 
explain how they are modified in a system with polar symmetry.

We assume that, apart from their activity, the assembly of particles can be well approximated as nematic liquid 
crystal --- that is, there is no difference between the two ends of an individual particle. The effect of relaxing 
this assumption by allowing the particles to be polar is discussed at the end of this letter. The hydrodynamics of 
nematic liquid crystals is well developed \cite{deGennesLCBook}. 
We adopt the Erickson-Leslie-Parodi approach \cite{VoiEPL05}, in which the nematic order parameter is a 
fixed-magnitude unit vector field $\vec{n}$ which evolves according to
\begin{equation}
\partial_t n_i + \vec{u}\cdot\vec{\nabla}n_i = \lambda D_{ij}n_j - \Omega_{ij} n_j + \Gamma h_i
\label{nEOM}
\end{equation}
where $\vec{u}$ is the velocity field of the solvent; $\lambda$ is the flow alignment parameter, more on which 
below; $\Gamma$ is a rotational viscosity; $\vec{h}$ is the molecular field given by
\begin{equation}
h_i = K\nabla^2n_i
\end{equation}
(assuming only a single elastic constant $K$); and
\begin{equation}
D_{ij} = {1\over2}(\partial_i u_j + \partial_j u_i) \ \ \  ; \ \ \ \Omega_{ij} = {1\over2}(\partial_i u_j - 
\partial_j u_i) .
\end{equation}
The first two terms on the right-hand side of   eq.~(\ref{nEOM}) describe alignment (or tumbling) of the director 
field by local shear flow. The third term accounts for the tendency of the ordered nematic to resist distortions, 
and arises ultimately from excluded volume interactions between individual particles. 

\begin{figure}
\includegraphics[width=3in]{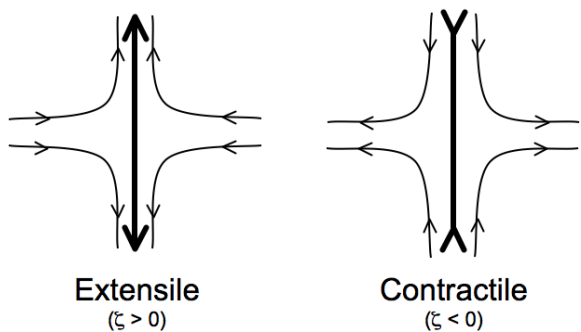}
\caption{The dipolar flow fields (curved arrows) generated by extensile (left) and contractile (right) particles. 
The straight vertical arrows represent the director field, which is along the rod axis for rod-like ($\lambda > 0$) 
particles and perpendicular to the disk plane for discoidal ($\lambda < 0$) particles.}
\label{zeta}
\end{figure}

The flow field $\vec{u}$ obeys the Navier-Stokes equation
\begin{eqnarray}
\rho(\partial_t+ \vec{u} \cdot \nabla)
u_i & =& \partial_j (\sigma_{ij})+
\eta \partial_j(\partial_i
u_j + \partial_j u_i)
\end{eqnarray}
where $\rho$ is the fluid density and $\eta$ is the viscosity. The passive part of the stress tensor $\sigma_{ij}$ 
is
\begin{equation}
\sigma^{p}_{ij} = -p \delta_{ij} 
- {\lambda \over 2} [n_i h_j + n_j h_i]
 + {1\over 2} [n_i h_j - n_j h_i] 
\end{equation}
where $p$ is the bulk pressure. 

Eqs.\ (1) --- (5) are generalised to describe active systems by introducing one or more terms $\sigma^{a}_{ij}$ to 
the stress tensor, so that $\sigma = \sigma^{p} + \sigma^{a}$. These active stresses account for the forces 
generated by the individual particles, as a function of the local director field, and cannot be derived from a free 
energy; however, their form can be arrived at from considering the symmetry of the flow field generated by the 
particles. It is commonly assumed that this flow field can be described to leading order as dipolar, and that 
higher order contributions can be neglected, when considering far-field interactions and long time scales 
\cite{SimPRL02}. The appropriate active stress tensor for a suspension of force dipoles is
\begin{equation}
\sigma^{a}_{ij} = -\zeta n_i n_j
\label{act_str}
\end{equation}
where $\zeta$ is the activity coefficient, the meaning of which is discussed below. One can also add active terms 
to the director equation of motion,   eq.~(\ref{nEOM}), to account for self-alignment effects \cite{VoiEPL05}, but 
as such terms do not play an important role in the spontaneous flow transition they are neglected here. For the 
same reason, we neglect active contributions to the isotropic pressure $p$.

The two key parameters that characterise the individual particles are the activity parameter $\zeta$ and the flow 
alignment parameter $\lambda$. Both the magnitude and sign of each have an easily understood physical significance. 
The sign of $\zeta$ determines whether the dipolar flow field generated by the particles is \emph{extensile} 
($\zeta > 0$) or \emph{contractile} ($\zeta <0$), as illustrated in fig.\  \ref{zeta}. In the swimmer literature, 
an alternative nomenclature is sometimes used, with extensile swimmers described as \emph{pushers} and contractile 
swimmers as \emph{pullers}. The magnitude of $\zeta$ relates to the strength of the forcing, or equivalently the 
rate at which the particles expend energy. 

The flow alignment parameter $\lambda$ determines how the director field responds to a shear flow. 
A shear flow can be decomposed into an extensional and a rotational component, and the magnitude of 
$\lambda$ determines their relative influence. For $|\lambda| < 1$, the rotational part of the flow 
always dominates regardless of the director orientation, and thus the director will continuously rotate 
under shear. This is the \emph{flow tumbling} regime. For $|\lambda| > 1$, however, the director will 
tend to align at a unique angle to the flow direction, $\theta_L = \frac{1}{2}\cos^{-1}\frac{1}{\lambda}$, at which 
the extensional and rotational parts of the shear flow balance. This is the \emph{flow aligning} regime. In 
general, molecule-sized nematic liquid crystals are flow aligning and larger objects are flow tumbling, since for 
the extensional flow to dominate there must be enough Brownian rotational diffusion to wash out the rotational part 
of the flow. The sign of $\lambda$ is also important, and relates to the shape of the individual particles. 
Rod-like objects have $\lambda > 0$; discoid objects have $\lambda < 0$; and $\lambda = 0$ corresponds to spheres. 

\begin{figure*}
\includegraphics[width=6.2in]{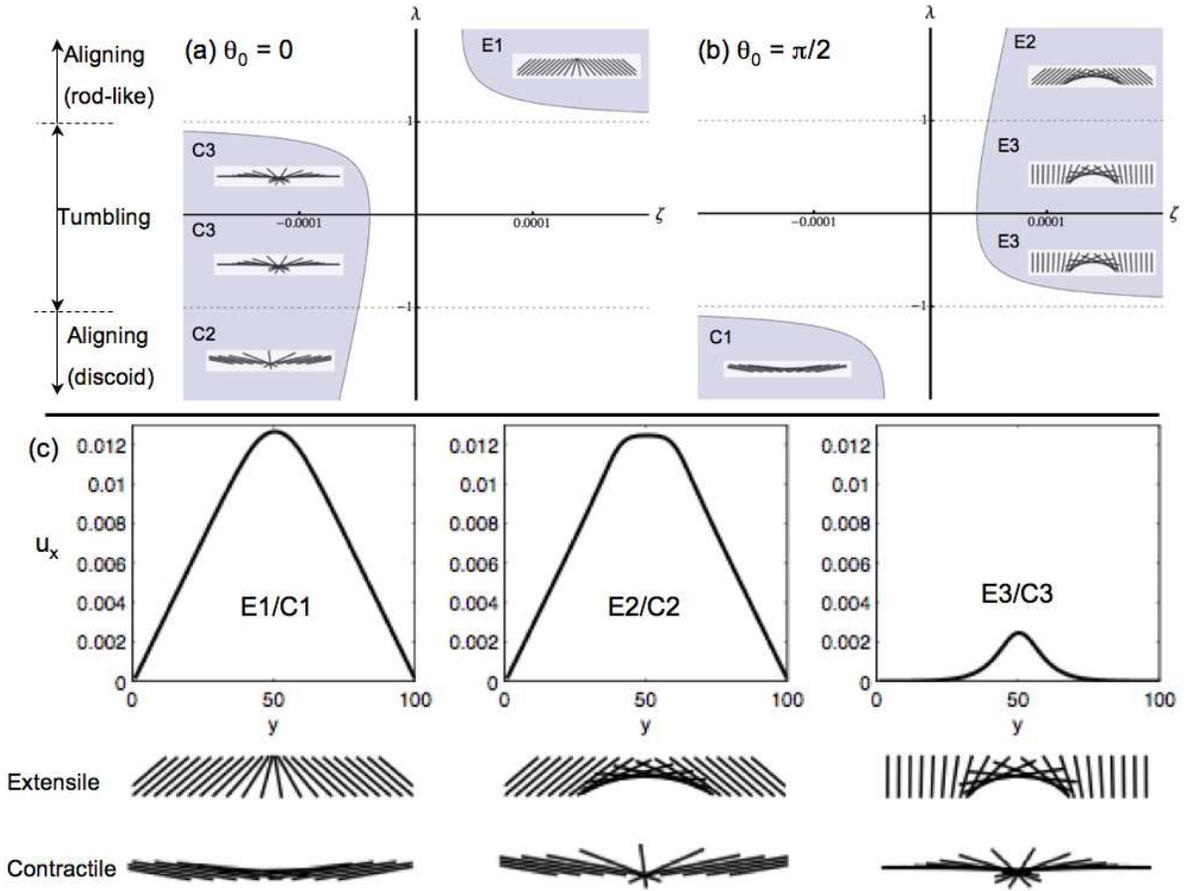}
\caption{Stable states (in activity $\zeta$ versus flow alignment parameter $\lambda$) for a one-dimensional 
slab of active nematic. Initial director orientation (a) parallel to the walls $\theta_0 = 0$. (b) perpendicular 
to the walls $\theta_0 = {\pi\over2}$. In the shaded (unshaded) regions the initial state is unstable (stable) to 
small perturbations. The phase boundaries are given by  eq.~(\ref{zetac}). The diagrams 
show the configuration of the director field $\vec{n}$ for the steady spontaneous flow state that arises in each 
unstable region. (c) Example flow fields $u_x(y)$ corresponding to the spontaneous flow states identified in (a) 
and (b). 
Parameters used are: for E1/C1, $\lambda =$ +1.5 / -1.5, $\zeta = $ +0.001/ -0.001, $\theta_0 =$ 0 / $\pi\over2$; 
for E2/C2, $\lambda =$ +1.5 / -1.5, $\zeta = $ +0.001/ -0.001, $\theta_0 = {\pi\over2}$ / 0; 
and for E3/C3, $\lambda =$ 0 / 0, $\zeta = $ +0.001/ -0.001, $\theta_0 = {\pi\over2}$ / 0. Other parameters are 
given in the text. }
\label{PDFF}
\end{figure*}

\begin{figure*}
\includegraphics[width=4.8in]{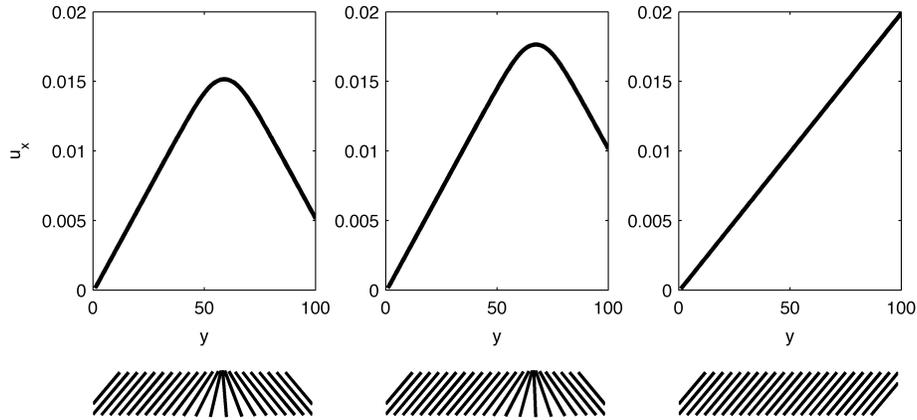}
\caption{Response of an E1-type kink to an applied shear, and the corresponding induced flow profile. The shear is 
generated by moving the right wall with a steady upwards velocity $v_{\mathrm{wall}}$. From left to right, 
$v_{\mathrm{wall}} = 0.005, 0.01, 0.02$ in simulation units.}
\label{E1shear}
\end{figure*}

The original analytic prediction of the spontaneous flow transition by Voiturez \emph{et al.}\ \cite{VoiEPL05} 
considered 
the contractile ($\zeta < 0$), discoidal, flow aligning ($\lambda < -1$) portion of phase space. Previous 
numerical studies by Marenduzzo \emph{et al.}\ \cite{MarPRE07} considered both extensile and contractile 
systems, for rod-like particles ($\lambda > 0$). We numerically explore the complete ($\zeta,\lambda$) 
phase space, and vary the orientation $\theta_0$ of the initial uniformly aligned state, 
to access a number of novel spontaneous flow states.

Eqs.\ (1)--(6) were solved, using a hybrid Lattice Boltzmann approach analogous to that described by 
Marenduzzo \emph{et al.} \cite{MarPRE07}, in the one dimensional geometry of Fig.\  \ref{slab}. We follow 
Cates \emph{et al.}\ \cite{CatArX08} in using free boundary conditions for the director 
field ($\partial_y n_i = 0$ at the walls); the effect of fixed boundary conditions will be mentioned 
below. The following parameter values were used throughout: slab width $L = 100$, $K = 0.04$, 
$\eta = 1.27$, and $\Gamma = 0.2$ (all in simulation units). The director field was initially 
uniform and aligned at an angle $\theta_0$ to the walls. To break the symmetry we perturbed the 
uniform state by generating a small kink at the centre of the slab by rotating the director in the 
left half of the domain clockwise 0.1 radians and that in the right half the same amount 
anti-clockwise. The diagrams in fig.\  \ref{PDFF} show where the system was found to be stable 
(unshaded) and unstable (shaded) against the perturbation, for (a) $\theta_0 = 0$, and (b) $\theta_0 = \pi/2$. 
The critical value of the activity $\zeta_c$ corresponding to the phase boundary is determined analytically by 
linearizing the equations of motion around the initial state, and checking when a non-uniform solution exists, 
following Voituriez \emph{et al.}\ \cite{VoiEPL05}
 \begin{equation}
\zeta_c = \left\{
 \begin{array}{rl}
	\frac{\pi^2 K (4 \eta \Gamma + (\lambda - 1)^2)}{2 L^2 (\lambda - 1)} & \textrm{for } \theta_0 = 0\\
	 & \\
	\frac{\pi^2 K (4 \eta \Gamma + (\lambda + 1)^2)}{2 L^2 (\lambda + 1)} & \textrm{for } \theta_0 = \frac{\pi}{2}.
 \end{array} \right.
 \label{zetac}
 \end{equation}
 
The director configurations in fig.~\ref{PDFF} are typical numerical results obtained after the spontaneous 
flow state is reached in each of the unstable regions. There are six in total, three each for extensile (labeled E1, E2, E3) and contractile (C1, C2, C3) systems. In the previous work by Marenduzzo \emph{et al.}\  \cite{MarPRE07,MarJNNFM08}, 
regions E1 and C3 were explored. The C2 state was predicted analytically in Voituriez \emph{et al.}\ 
\cite{VoiEPL05} but has not yet been studied numerically. 

The six kink states are, for the geometry we consider, related in pairs by a symmetry transformation.
This symmetry is apparent from an examination of the flow profiles generated by each type of spontaneous 
flow state, shown in fig.\  \ref{PDFF}(c). Notice that the states naturally fall into three pairs 
(E1/C1, E2/C2 and E3/C3), each with the same type of flow profile, but director fields rotated by 
$\pi / 2$ radians with respect to each other. The states in each pair have $\zeta$ and $\lambda$ 
values with the same magnitudes, but opposite signs. This highlights the symmetry inherent 
in the equations of motion: a change in sign of $\zeta$ is equivalent to a change in sign of 
$\lambda$ together with a $\pi / 2$ rotation of the director field; an extensile rod-like particle 
interacts with a flow field the same way as a contractile discoidal particle rotated by $\pi / 2$. 
Note that although this holds true for the quasi-one dimensional geometry of interest, it is not 
universally true; in particular the symmetry does not hold in three dimensions.

\begin{figure*}
\includegraphics[width=5.5in]{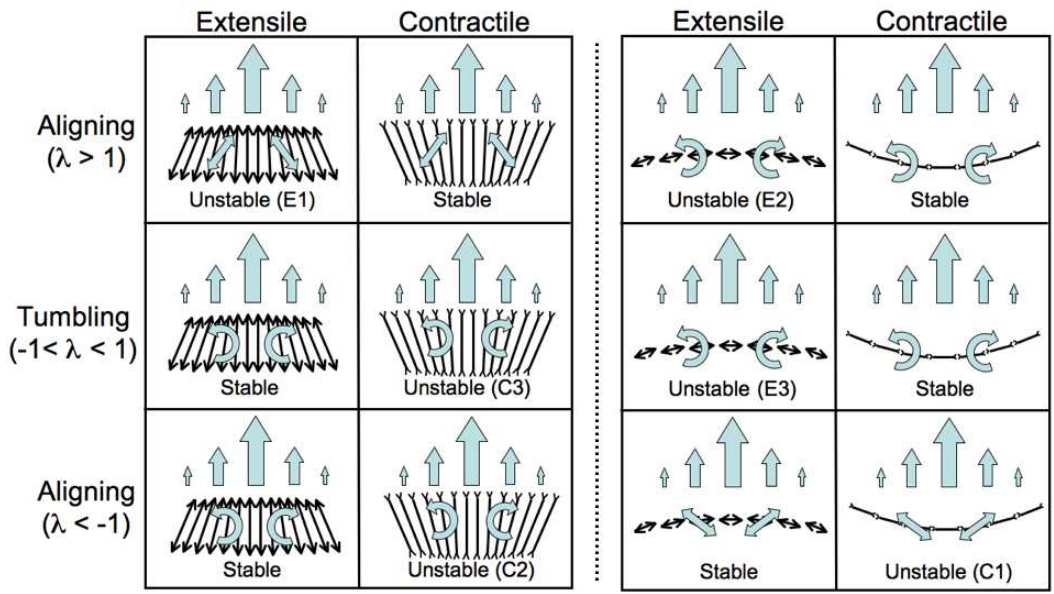}
\caption{Mechanism driving the (in)stability of an initially uniformly aligned state. Thin straight arrows show the 
director field, following a kink perturbation acting on a state with initial orientation $\theta_0 = 0$ (left) 
and $\theta_0 = \pi/2$ (right). Arrowheads pointing outwards (inwards) denote extensile (contractile). Wide 
vertical 
arrows show shear flow generated by the kink. Wide arrows superimposed on the director field show the influence of 
the shear flow upon it: either aligning (straight, double-headed arrows) or tumbling (curved arrows).}
\label{splaybend}
\end{figure*}

The instabilities leading to each type of kink can be understood from a simple pictorial model, extending 
ideas from \cite{RamNJP07} (who consider the flow tumbling regime). Fig.\  \ref{splaybend} illustrates kink 
perturbations to the initial state of the director field, $\theta_0  = 0 $ on the left and $\theta_0 = \pi/2$ 
on the right. The discontinuity in $\vec{n}$ at the centre of the kink generates a force on the fluid due 
to the active term in the stress tensor $\sigma^{p}$. The flow direction depends on the sign of $\zeta$; 
in the figure, all kinks are drawn such that the resulting flow is upwards. Because of the no-slip condition 
at the walls, this central forcing necessarily generates a region of uniform shear flow on either side of the 
kink, illustrated by the fat vertical arrows. The wide arrows overlaid on the director field show the effect 
of the shear flow on the director in each case: depending on the values of $\lambda$ and $\theta_0$ either 
the extensional flow dominates, attempting to align the director at the Leslie angle (double headed arrows) 
or the rotational flow dominates (curved arrows) and causes the director to tumble. By reacting thus to the 
shear flow, the director field either returns to the initial aligned state and the induced shear flow 
vanishes, or the perturbation grows until one of the stable spontaneous flow states is reached. The same 
instability mechanism applies in two dimensions, except that without walls to restrict flow to one axis 
there will be no steady spontaneous flow states. Instead, the instabilities drive quasi-turbulent mixing 
flow \cite{AraPRE07,SaiPRL08}.

The six stable kink states in fig.\  \ref{PDFF} are the fundamental building blocks for more complicated states 
observed in other numerical studies. For instance, if fixed boundary conditions are used for the director field as 
in Marenduzzo \emph{et al.}\ \cite{MarPRE07}, a half-kink is necessary at each wall. At higher activities, striped 
states can be observed, which are simply multiple kinks sitting side by side. We do not observe striped states in 
our study, even at higher $|\zeta|$ values, because our method for perturbing the initial state seeds a single kink 
at the centre. In contrast, Marenduzzo \emph{et al.}\ perturb by increasing the director orientation slightly at a 
single point, which stimulates the formation of at least two kinks, one on either side of the perturbation. Thus 
the exact state chosen by the system can depend strongly on how the initial uniform state is perturbed. It is 
possible to have the two different types of flow aligning kink (E1 and E2, or C1 and C2) coexisting in the same 
state, but only if the state is specially prepared. Note that we restrict the director to the $x$--$y$ plane, which 
does not allow for twisted states, as were sometimes observed for high $|\zeta|$ in \cite{MarPRE07}. 

Each of the states reacts differently to externally imposed shear. For the tumbling states, E3 and C3, 
even a small amount of imposed shear causes the director to rotate continuously, and there is no steady state. 
Indeed 
these states are only stable in the first place because they spontaneously adopt a configuration which generates 
no fluid flow, and thus no shear, outside the kink. 
In contrast, the flow aligning states E1/C1 and E2/C2 can survive in applied shear up to a certain 
magnitude. A single kink of either type responds to applied shear by moving towards the wall that is moving 
in the same direction as the spontaneous flow. This is illustrated in fig.\ \ref{E1shear} for an E1 kink. At 
some critical wall velocity, the kink gets too close to the wall and becomes unstable; for higher velocities, 
there is no steady kink state and the velocity profile is standard linear shear flow. 

Having catalogued this variety of states, it is important to ask which we would predict to occur in specific 
physical systems. Considering first swimming bacteria: they may be either extensile or contractile, depending on 
their mode of swimming, but most are certainly rod-like ($\lambda >0$) and furthermore should be flow tumbling 
($|\lambda| < 1$) given their relatively large size. Therefore, with reference to fig.\ \ref{PDFF}, one might 
expect to see states of type E3 for suspensions of ``pushers" (e.g.\ \emph{E.\ coli}) or C3 for ``pullers" (e.g.\  
\emph{Chlamydomonas}). This is consistent with the experiments and analytical predictions of Berke \emph{et al.}\ 
\cite{BerPRL08} showing that pushers tend to align parallel to walls, as seen in state E3, while pullers align 
perpendicularly as in C3. Note however that the existence of spontaneous flow in these states relies on a non-zero elastic 
constant $K$; if $K \rightarrow 0$, the width of the kink also goes to zero, and for the states E3 and C3 the 
induced flow would vanish. 
Therefore attempts to simulate suspensions of microswimmers must incorporate liquid crystalline elasticity, for 
example via excluded volume interactions, to see steady flow-generating kinks of this type. Previous simulations of 
large numbers of
rod-like swimmers in the literature \cite{HerPRL05, SaiPRL07} do not incorporate excluded volume interactions, and 
therefore would not be expected to reproduce any such steady spontaneous flow states. However, our predictions are 
consistent with the instabilities of uniform aligned states observed in these simulations. In particular, the 
enhancement  of mixing by pushers (extensile rods) but not by pullers  (contractile rods) \cite{UndPRL08,SaiPRL08} 
can be explained with reference to fig.\ \ref{splaybend}. For the value $\lambda = 1$ assumed by 
Saintillan \emph{et al.}\ \cite{SaiPRL07}, an aligned state of contractile rods is stable whereas one 
of extensile rods is unstable.

For solutions of cytoskeletal rods and molecular motors there is experimental evidence for the active stress being 
contractile \cite{ThoJCellSci97}, but it is not clear whether they should be treated as flow aligning or flow 
tumbling. Voituriez \emph{et al.}\ \cite{VoiEPL05} assumed the former, in which case the particles must be 
discoidal ($\lambda < -1$) for spontaneous flow to occur, via states C1 or C2. If they are flow tumbling, then they 
could be expected to generate kinks of type C3. In either case, our results suggest that cytoskeletal components 
will tend to align roughly perpendicular to the walls, if the director is not otherwise pinned at the boundary.  

\begin{figure}
\includegraphics[width=3.0in]{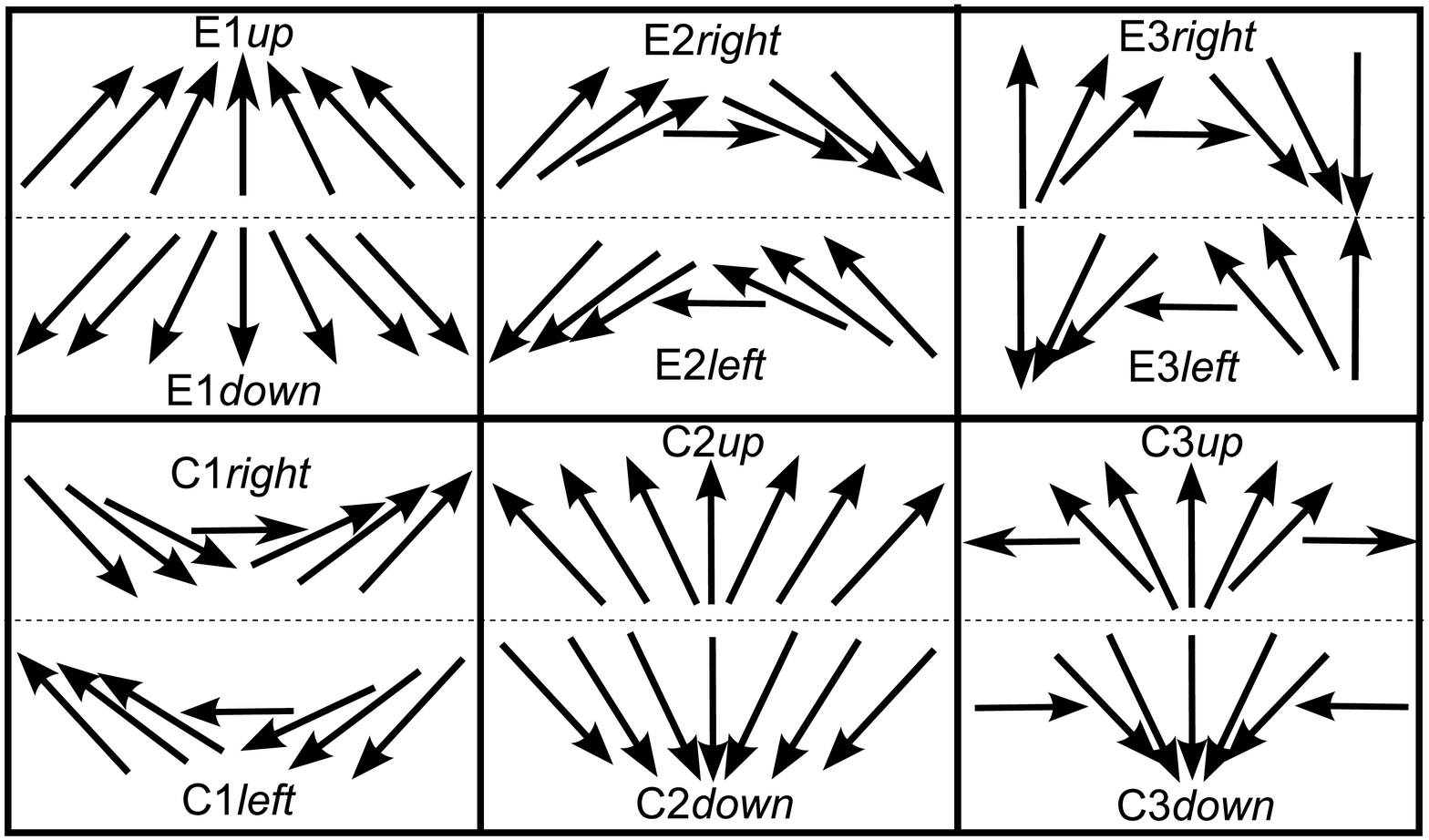}
\caption{The twelve distinct kink types that can occur when the particles are polar. Only six of them --- those 
with `up/down' alignment --- are persistent long-term states, with the others moving sideways after formation until 
they are annihilated at a wall. }
\label{polar}
\end{figure}

The story changes significantly if the particles are allowed to be polar, having a preferred direction along which 
they will move. Polarity is accounted for in the model by including more active terms that are forbidden by 
symmetry in the nematic case. At lowest order, these comprise an extra active contribution to the stress tensor 
$\sigma^{\mathrm{pol}}_{ij} = \beta(\partial_in_j + \partial_jn_i)$, and the replacement of $\vec{u}$ in the 
director equation of motion (eq.\ \ref{nEOM}) with $\vec{u}+\beta\vec{n}$ (for simplicity we assume that both these 
terms carry the same polar activity coefficient $\beta$, but in general they may be different). When $\beta$ is 
allowed to be non-zero, the number of distinct kink states is doubled with respect to the nematic case (fig.\ 
\ref{polar}) since each of the original kinks illustrated in fig.\ \ref{PDFF} now has two possible alignments. The 
nematic states for which the kink centre is parallel to the walls (E1, C2 and C3) each have `up' and `down' polar 
analogues; the nematic kinks with centres perpendicular to the walls (C1, E2 and E3) have `left' and `right' 
analogues. Numerical studies of the polar kinks show that only the `up/down' kinks remain stationary and 
persistent. The `left/right' kinks, by contrast, move in the direction they point until they reach a wall, and then 
vanish, leaving a uniformly aligned state with zero induced flow. Note that the symmetry between the three pairs of states E1/C1, E2/C2 and E3/C3 is broken by the introduction of polarity. The remaining stationary
kinks can be further subdivided into two groups: those that point inwards (E1\emph{up}, C2\emph{down} and C3\emph{down}), for which the sharpness of the
peak in the induced velocity profile increases with $\beta$, and those that point outwards (E1\emph{down}, C2\emph{up} and C3\emph{up}) for which it decreases.
If the local
concentration of particles is allowed to vary, evolving via diffusion and self-advection, the particle concentration is enhanced at the centre of the kink and reduced at the walls for the inward kinks, but tends to build up at the walls for the outward kinks.

We have unified and extended previous studies of spontaneous flow in quasi-1D active nematics by numerically exploring 
the range of possible steady states as the activity, flow alignment characteristics and orientation are varied. 
We predict three qualitatively 
distinct types of spontaneous flow state, and examine how they respond to externally applied shear. We have also 
studied how the ensemble of stable states is affected when the active particles are made polar. 
Knowledge of the range of possible states should help match continuum and microscopic models of active materials 
and guide future experiments and simulations attempting to realize spontaneous fluid flow in narrow channels. \\
~\\
\acknowledgments
We thank G. Alexander, C. M. Marchetti, M. E. Cates and E. Orlandini for helpful discussions and acknowledge financial support from 
ONR and EPSRC.


\begin{thebibliography}{10}
\expandafter\ifx\csname url\endcsname\relax\def\url#1{\texttt{#1}}\fi

\bibitem{VoiEPL05}
Voituriez R., Joanny J.~F. and Prost J. {\it Europhys. Lett.} {\bf 70}, 404 (2005).

\bibitem{DomPRL04}
Dombrowski C., Cisneros L., Chatkaew S., Goldstein R.~E. and Kessler
  J.~O. {\it Phys. Rev. Lett.} {\bf 93}, 098103 (2004).

\bibitem{NedNature97}
Nedelec F.~J., Surrey T., Maggs A.~C. and Leibler S. {\it Nature} {\bf 389}, 305 (1997).

\bibitem{NedCurrOpin03}
Nedelec F., Surrey T. and Karsenti E. {\it Curr. Opin. Cell Biol.} {\bf 15}, 118 (2003).

\bibitem{SimPRL02}
Simha A.~R. and Ramaswamy S. {\it Phys. Rev. Lett.} {\bf 89}, 058101 (2002).

\bibitem{HatPRL04}
Hatwalne Y., Ramaswamy S., Rao M. and Simha A.~R. {\it Phys. Rev.
  Lett.} {\bf 92}, 118101 (2004).

\bibitem{AraPRE07}
Aranson I.~S., Sokolov A., Kessler J.~O. and Goldstein R.~E.
  {\it Phys. Rev. E} {\bf 75}, 040901 (2007).

\bibitem{SaiPRL08}
Saintillan D. and Shelley M.~J. {\it Phys. Rev. Lett.} {\bf 100}, 178103 (2008).

\bibitem{LivPRL03}
Liverpool T.~B. and Marchetti C.~M. {\it Phys. Rev. Lett.} {\bf 90}, 138102 (2003).

\bibitem{LivPRL06}
Liverpool T.~B. and Marchetti C.~M. {\it Phys. Rev. Lett.} {\bf 97}, 268101 (2006).

\bibitem{KruPRL04}
Kruse K., Joanny J.~F., J\"ulicher F., Prost J. and Sekimoto K.
  {\it Phys. Rev. Lett.} {\bf 92}, 078101 (2004).

\bibitem{KruEPJE05}
Kruse K., Joanny J.~F., J\"{u}licher F., Prost J. and Sekimoto K.
  {\it Euro. Phys. J. E} {\bf 16}, 5 (2005).

\bibitem{VoiPRL06}
Voituriez R., Joanny J.~F. and Prost J. {\it Phys. Rev. Lett.} {\bf 96}, 028102 (2006).

\bibitem{MarPRL07}
Marenduzzo D., Orlandini  E. and Yeomans J.~M. {\it Phys. Rev. Lett.} {\bf 98}, 118102 (2007).

\bibitem{MarPRE07}
Marenduzzo D., Orlandini E., Cates M.~E. and Yeomans J.~M.
  {\it Phys. Rev. E} {\bf 76}, 031921(2007).

\bibitem{MarJNNFM08}
Marenduzzo D., Orlandini E., Cates M.~E. and Yeomans J.~M. {\it J.
  Non-Newt. Fluid Mech.} {\bf 149}, 56 (2008).

\bibitem{deGennesLCBook}
de~Gennes P.~G. and Prost J. {\it The Physics of Liquid Crystals}
  (Oxford University Press) 1995.

\bibitem{CatArX08}
Cates M.~E., Fielding S.~M., Marenduzzo D., Orlandini E. and Yeomans
  J.~M. {\it Phys. Rev. Lett.} {\bf 101}, 068102 (2008).

\bibitem{RamNJP07}
Ramaswamy S. and Rao M. {\it New J. Phys.} {\bf 9}, 423 (2007).

\bibitem{BerPRL08}
Berke A.~P., Turner L., Berg H.~C. and Lauga E. {\it Phys. Rev.
  Lett.} {\bf 101}, 038102 (2008).

\bibitem{HerPRL05}
Hernandez-Ortiz J.~P., Stoltz C.~G. and Graham M.~D. {\it Phys. Rev.
  Lett.} {\bf 95}, 204501 (2005).

\bibitem{SaiPRL07}
Saintillan D. and Shelley M.~J. {\it Phys. Rev. Lett.} {\bf 99}, 058102 (2007).

\bibitem{UndPRL08}
Underhill P.~T., Hernandez-Ortiz J.~P. and Graham M.~D. {\it Phys.
  Rev. Lett.} {\bf 100}, 248101 (2008).

\bibitem{ThoJCellSci97}
Thoumine O. and Ott A. {\it J. Cell Sci.} {\bf 110}, 2109 (1997).

\end{thebibliography}
\end{document}